\documentclass[10pt]{article}

\usepackage{amsmath}

\usepackage{array}

\usepackage{appendix}

\usepackage{tocloft}                   

\usepackage{graphicx}

\usepackage{amsfonts}

\usepackage{amssymb}

\usepackage{mathrsfs}

\usepackage{yfonts}

\usepackage{euscript}

\usepackage{centernot}                 

\usepackage{ifsym}                     

\usepackage{lmodern}                   

\usepackage{upgreek}

\usepackage{mathtools}

\usepackage{color}

\usepackage{slantsc}
\usepackage{calligra}

\usepackage{bbold}          

\usepackage[T1]{fontenc}

\usepackage{epsf}

\usepackage{latexsym}

\usepackage{tipa}

\usepackage{makeidx}

\makeindex

\def\b{\begin{equation}}

\def\e{\begin{equation}}

\def\be{\begin{equation}}              

\def\ee{\end{equation}}

\def\beq{\begin{equation}}

\def\eeq{\end{equation}}

\def\bea{\begin{eqnarray}}

\def\eea{\end{eqnarray}}

\def\m{\mbox{ }}

\def\!{\hspace{-1.6667em}}

\def\n{\noindent}

\def\u{\underline}

\def\bia{\mbox{\boldmath$a$}}

\def\biA{\mbox{\boldmath$A$}}

\def\biP{\mbox{\boldmath$P$}}

\def\biQ{\mbox{\boldmath$Q$}}

\def\biR{\mbox{\boldmath$R$}}

\def\biS{\mbox{\boldmath$S$}}

\def\biV{\mbox{\boldmath$V$}}

\def\sbiO{\mbox{\ttfamily\fontseries{b}\selectfont O}} 

\def\sbiV{\mbox{\scriptsize\boldmath$V$}}

\def\sbiA{\mbox{\scriptsize\boldmath$A$}}

\def\bLambda{\mbox{\boldmath$\Lambda$}}             

\def\mC{\mbox{C}}                        

\def\mH{\mbox{H}}

\def\ba{\mbox{\bf a}}

\def\bsigma{\mbox{\boldmath$\sigma$}}                   %

\def\sf{\mbox{\scriptsize f}}

\def\si{\mbox{\scriptsize i}}

\def\sn{\mbox{\scriptsize n}}

\def\sG{\mbox{\scriptsize G}}

\def\sN{\mbox{\scriptsize N}}

\def\sS{\mbox{\scriptsize S}}

\def\sV{\mbox{\scriptsize V}}

\def\sbcC{\mbox{\boldmath \scriptsize ${\cal C}$}}

\def\sbcG{\mbox{\boldmath \scriptsize ${\cal G}$}}

\def\bscF{\mbox{{\boldmath \scriptsize${\cal F}$}}}                               

\def\sumi2{\sum\mbox{}_{\mbox{}_{\mbox{\scriptsize $i$=1}}}^2}

\def\sumi3{\sum\mbox{}_{\mbox{}_{\mbox{\scriptsize $i$=1}}}^3}

\def\sumABcycles3{\sum\mbox{}_{\mbox{}_{\mbox{\scriptsize cycles $A,B$=1}}}^{3}}

\def\sumCDcycles3{\sum\mbox{}_{\mbox{}_{\mbox{\scriptsize cycles $C,D$=1}}}^{3}}

\def\sumj3{\sum\mbox{}_{\mbox{}_{\mbox{\scriptsize $j$=1}}}^3}

\def\sumk3{\sum\mbox{}_{\mbox{}_{\mbox{\scriptsize $k$=1}}}^3}

\def\prodiA1{\prod\mbox{}_{\mbox{}_{\mbox{\scriptsize $i$=1}}}^{A - 1}}

\def\d{\textrm{d}}                                                  

\def\es{\m = \m}

\def\:={\m := \m}

\def\=:{\m =: \m}

\def\peq{\m \mbox{`='} \m}

\def\FrX{\mathfrak{X}}                                         

\def\lFrg{\mbox{\Large$\mathfrak{g}$}}                         
\def\Frg{\mbox{\normalsize $\mathfrak{g}$}}                    

\def\bFrV{\mbox{\boldmath$\mathfrak{V}$}}                      

\def\Hilb{\mbox{{\boldmath$\mathfrak{H}$}ilb}}                 

\def\scG{\mbox{\scriptsize ${\cal G}$}}

\def\scG{\mbox{\scriptsize ${\cal G}$}}                    

\def\Phase{\mbox{{\boldmath$\mathfrak{P}$}hase}}                     

\def\bFrR{\mbox{\boldmath$\mathfrak{R}$}}                            
\def\Rig-Phase{\bFrR\mbox{ig-}\Phase}                                
                                                                													   
\def\bFrR{\mbox{\boldmath$\mathfrak{R}$}}                            

\def\bFrR{\mbox{\boldmath$\mathfrak{R}$}}                            

\def\1mat{\u{\u{1}}}                                                 

\def\Positive-Modespace{\mbox{{\boldmath$\mathfrak{M}$}odespace$^+$}}

\def\POSITIVE-MODESPACE{\mbox{{\boldmath$\mathfrak{M}$}ODESPACE$^+$}}
                                                                                                                             														
\def\Obs{\FrO\mbox{bs}}

\def\FrO{\mbox{$\mathfrak{O}$}}                                      

\def\Kin-Hilb{\mbox{{\boldmath$\mathfrak{K}$}in-\Hilb}}                     

\def\Mid-Hilb{\mbox{{\boldmath$\mathfrak{M}$}id-\Hilb}}                     

\def\Dyn-Hilb{\mbox{{\boldmath$\mathfrak{D}$}yn-\Hilb}}                     

\def\5Star{\mbox{\Large$\star$}}              

\begin{document}

\begin{center}

\Large{\bf Nijenhuis-type variants of Local Theory of Background Independence}\normalsize

\vspace{0.1in}

{\large \bf Edward Anderson}$^1$ 

\end{center}

\begin{abstract}

A local resolution of the Problem of Time has recently been given, alongside reformulation as a local theory of Background Independence.   
The classical part of this can be viewed as requiring just Lie's Mathematics, 
albeit entrenched in subsequent Topology and Differential Geometry developments and extended to the setting of contemporary Physics' state spaces.
We now generalize this approach by mild recategorization to one based on Nijenhuis' generalization of Lie's Mathematics, as follows. 
1)  Relationalism is encoded using the Nijenhuis--Lie derivative. 
2) Closure is assessed using the Schouten--Nijenhuis bracket, and a `Schouten--Nijenhuis Algorithm' analogue of the Dirac and Lie Algorithms.  
This produces a class of Gerstenhaber algebraic structures of generators or of constraints.   
3) Observables are defined by a Schouten--Nijenhuis brackets relation, 
   reformulating the constrained canonical case as explicit PDEs to be solved using the Flow Method, 
   and forming their own Gerstenhaber algebras of observables. 
Lattices of Schouten--Nijenhuis--Gerstenhaber constraint or generator algebraic substructures 
furthermore induce dual lattices of Gerstenhaber observables subalgebras.  
4) Deformation of Gerstenhaber algebraic structures of generators or constraints encountering Rigidity gives a means of Constructing more structure from less.    
5) Reallocation of Intermediary-Object Invariance gives the general Schouten--Nijenhuis--Gerstenhaber algebraic structure's 
analogue of posing Refoliation Invariance for GR. 
We finally point to general Gerstenhaber bracket and Vinogradov bracket generalizations, 
with the former likely to play a significant role in Backgound-Independent Deformation Quantization and Quantum Operator Algebras.  

\end{abstract}

$^1$ dr.e.anderson.maths.physics *at* protonmail.com

\section{Introduction}\label{Introduction}

It has been recently demonstrated that \cite{ABook, ALRoPoT, Higher-Lie, XIV} Lie's Mathematics suffices to construct A Local Resolution 
of the Problem of Time \cite{Battelle-DeWitt67, Dirac, K92-I93, APoT-2, ABook}, 
which in turn can be reformulated as \cite{ABook, A-CBI} A Local Theory of Background Independence.

This locally-smooth approach is moreover sufficiently well-defined to extend to various other (at least locally) differential-geometric structures. 
The purpose of the current Article is to outline one of the more interesting cases: the `Nijenhuis Mathematics' \cite{S40-53-N55, FN56, NR66} counterpart; 
see \cite{Nambu} for use of `Nambu Mathematics' instead, while \cite{ABook, XIV} already made mention of the simpler graded, alias supersymmetric, extension.    
Nijenhuis Mathematics' distinctive primary structures are outlined in Sec 2, 
with further generalizations outlined in the concluding Sec 4: 
Vinogradov brackets \cite{V90, KS} -- a unification of Nijenhuis brackets --  
and general Gerstenhaber algebras.
The latter motivates our current study, due to its Deformation Quantization \cite{L78, Landsman, Kontsevich, Gengoux} and 
                                                   quantum operator algebra applications.

The main part of the current Article is Sec 3, where the Abstract's structures 1) to 5) 
-- Nijenhuis parallels of the Lie structures used in A Local Resolution of the Problem of Time and its reformulation as A Local Theory of Background Independence -- 
are outlined. 
These include in particular a `Nijenhuis Algorithm' analogue of the Dirac Algorithm, and a more general theory of observables than
those based on, sequentially, Dirac's Mathematics \cite{DiracObs, Dirac, HTBook, ABook} or Lie's \cite{AObs, XIV}.

\section{Nijenhuis Mathematics}\label{NM}

\n{\bf Definition 1} The {\it Schouten--Nijenhuis (SN) bracket} \cite{S40-53-N55, Gengoux}\footnote{Also note some related types of bracket are in use: 
Fr\"{o}licher--Nijenhuis bracket\cite{FN56, Michor} and Nijenhuis--Richardson bracket\cite{NR66}, 
each on vector-valued differential forms rather than on multivector fields.} 
on degree-r and thus shifted degree $\bar{r} := r - 1$ multivector fields $\FrX^r$ is given by
$$ 
\mbox{\bf [} \m \mbox{\bf ,} \, \m \mbox{\bf ]}_{\sS\sN} : \FrX^{\bar{p}} \times \FrX^{\bar{q}} \longrightarrow \FrX^{\bar{p} + \bar{q}}
\label{NS-Bracket}
$$ 
$$ 
\mbox{\bf [} \, \biP \mbox{\bf ,} \, \biQ  \, \mbox{\bf ]}_{\sS\sN}  (F_1, \, ... \, , \, F_{\bar{p} + \bar{q} + 1} ) \:= \m \m \m \m \m \m \m \m \m \m \m \m  
                        \sum_{  \sigma \in S_{q, \bar{p}}  } \, \mbox{sign}(\sigma) 
\biP ( \biQ ( F_{\sigma(1)}, \, ... \, F_{\sigma(q)}) , \, F_{\sigma(q + 1)} , \, ... \, F_{\sigma(q + \bar{p})} )      \m - \m 
$$
\be
\hspace{2.3in} -(-1)^{\bar{p}\bar{q}}  \sum_{  \sigma \in S_{p, \bar{q}}  } \, \mbox{sign}(\sigma) 
\biQ ( \biP ( F_{\sigma(1)}, \, ... \, F_{\sigma(p)}) , \, F_{\sigma(p + 1)} , \, ... \, F_{\sigma(p + \bar{q})} )        \m \m \m \m \m \m \m \m ,  
\ee 
where $\sigma$ denotes a shuffle and $S$ a permutation group formed by such. 
This obeys 
\be 
                       \mbox{\bf [} \, \biP \mbox{\bf ,} \, \biQ \, \mbox{\bf ]}_{\sS\sN}                         \es  
-(-1)^{\bar{p}\bar{q}} \mbox{\bf [} \, \biQ \mbox{\bf ,} \, \biP \, \mbox{\bf ]}_{\sS\sN}  \m \m \m \m \m \m \m \m \m \m \m \mbox{ (graded antisymmetry)}       \m , 
\ee  
\be 
(-1)^{\bar{p}\bar{r}} \mbox{\bf [} \, \biP \mbox{\bf ,} \, \mbox{\bf [} \, \biQ \mbox{\bf ,} \,  \biR \, \mbox{\bf ]}_{\sS\sN} \, \mbox{\bf ]}_{\sS\sN}          \m + \m 
\mbox{cycles } \es 0                                                                 \m \m \m                         \mbox{ (graded Jacobi identity)}    \m . 
\ee 
\vspace{1in}

\n{\bf Definition 2} The {\it Nijenhuis--Lie derivative} of $\biP  \in  \FrX^p$  with respect to  $\biV \in \FrX^1$  is 

\n\be 
\pounds^{\sN}_{\sbiV} \biP  \:=  \mbox{\bf [} \, \biV \mbox{\bf ,} \, \biP \, \mbox{\bf ]}_{\sS\sN}                                                          \m . 
\label{NL-Deriv}
\ee 
\n{\bf Structure 3} The algebras formed by equipping (graded) vector spaces $\bFrV$ with SN brackets are a subcase of {\it Gerstenhaber algebras} (defined in Sec 4),  
so we refer to them as {\it SNG-algebras}.
Let us finally also extend consideration from algebras to algebroids \cite{CM}, using the phrase `algebraic structures' as a portmanteau of the two. 
Each of classical deformation theory \cite{CM, Higher-Lie}, kinematical quantization \cite{Landsman}, 
and GR producing a constraint algebroid -- the Dirac algebroid \cite{Dirac} -- even before either of the previous are involved, 
justify this more generalized scope.  
{\it SNG-algebroids} and {\it SNG-algebraic structures} are thus in play.

\section{Nijenhuis Local Background Independence}\label{NLBI}

\n 1) We here employ {\sl Nijenhuis--Lie derivatives} (\ref{NL-Deriv}) to encode Relationalism.   

\n A) In the canonical case, we work with changes of configuration 

\n\be 
\d \biQ
\ee 
in place of velocities $\dot{\biQ} = \d \biQ/d t$ so as to stay free from time variables for the reasons given in Article I of \cite{ALRoPoT}. 

\n B) We correct by Nijenhuis--Lie derivative along physically irrelevant group $\lFrg$'s changes $\d \ba$, 

\n\be 
\d \biQ \m \longrightarrow \m \d \biQ - \pounds^{\sN}_{\d \bia} \biQ  \m.  
\ee
\n C) We know what form these corrections take by solving the {\sl generalized Killing--Nijenhuis equation} 

\n\be 
\pounds^{\sN} \bsigma = 0
\ee 
for geometrical level of structure $\bsigma$ to obtain the corresponding physically irrelevant automorphism group in question, $\lFrg$. 

\n D) We complete this with a move using all of $\lFrg$ to obtain $\lFrg$-invariant objects.
[C.f.\ {\sl group averaging} or Article II of \cite{ALRoPoT} for a detailed review of all of B) to D)].

\n In the spacetime counterpart,\footnote{We consider both canonical and spacetime settings for the structures discussed in the current Article; 
see e.g. \cite{ABook, ALRoPoT} for discussions of the merit of each of these positions, as well as for inter-relations between these two positions.} 
we are free to use plain auxiliary corrections $\biA$ in place of change corrections $\d \bia$ on spacetime objects $\biS$: 

\n\be 
\biS \longrightarrow \biS - \pounds^{\sN}_{\sbiA} \biS  \m ,
\ee 
with steps B) and C) then applying unaltered.

\m 

\n 2) Closure is assessed using a) the SN bracket (\ref{NS-Bracket}).  

\n b) An `SN Algorithm' analogue of the Dirac \cite{Dirac} and Lie \cite{Lie, XIV} Algorithms.  
This permits six types of equation to arise from an initial set of generators $\sbcG$ or constraints $\sbcC$, as follows. 

\m 

\n i) {\it Inconsistencies}: equations reducing to $0 = 1$ as envisaged by Dirac \cite{Dirac}. 

\n ii) {\it Identities}: equations reducing to $0 = 0$. 

\n iii) {\it New secondary generators} $\sbcG^{\prime}$ or secondary constraints $\sbcC^{\prime}$.

\n iv) `{\it SN specifier equations}' are also possible if there is an appending process. 
I.e.\ a generalization of Dirac's appending of constraints to Hamiltonians $H$ using Lagrange multipliers $\bLambda$, i.e.\ 

\n\be 
H \longrightarrow H + \bLambda \cdot \sbcC  \m ,
\ee 
by which equations relating these a priori for $\bLambda$ can also arise from one's algorithm.  

\n v) {\it Rebracketing} using {\it SN--Dirac brackets} in the event of encountering {\it SN-secondary objects}, 
i.e.\ ones that do not close under SN brackets.

\n vi) {\it Topological obsruction terms}: Nijenhuis Mathematics' analogue of anomalies. 

\m 

\n {\sl Termination occurs} if the SN Algorithm (using \cite{XIV}'s terminology) 0)   {\it hits an inconsistency}, 
                                                                              I)   {\it cascades to inconsistency}, 
         												                      II)  {\it cascades to triviality}, or 
												                              III) {\it arrives at an iteration producing no new objects} 
																			            while retaining some degrees of freedom.   
\n Successful candidates -- terminating at iii) -- produce an `SNG' class of Gerstenhaber algebraic structures of generators 

\n\be
\lFrg^{\sS\sN\sG}  \m ,
\ee  
or of SN-first-class constraints, $\bscF$ (those that do close under SN brackets) in the canonical case.   
   
\m 

\n 3) We next consider {\it observables} $\sbiO$ \cite{AObs, DiracObs, K92-I93, ABook} 
defined in the present context by {\it zero-commutant SN brackets} with the generators, $\sbcG$ 
(including with the $\bscF$ constraints as a subcase)   

\n\be 
\mbox{\bf [} \, \sbcG, \, \sbiO \, \mbox{\bf ]}_{\sS\sN}  \peq 0   \m . 
\label{NS-Commutants}
\ee  
$\peq$ is here the portmanteau of Dirac's notion of weak equality $\approx$ \cite{Dirac} 
-- from equality up to a linear combination of constraints to \cite{XIV} equality up to a linear combination of generators -- 
and of strong vanishing: the standard notion of equality, =.
In the constrained canonical subcontext, these are zero-commutants of the SN-first-class constraints. 
In this context, these equations can moreover be recast as explicit PDEs to be solved using the Flow Method \cite{John, Lee2}.  
This particular application gives an {\it SN integral theory of invariants}.  
I.e.\ the SN version of {\it Lie's Integral Approach to Invariants} \cite{Lie, XIV},  
albeit in both cases elevated to restricting functional dependence on a function space over the phase space geometry in question.  
The spacetime version is, rather, over the space of spacetimes. 
Such functions are, in each case, observables.  
These functions moreover constitute SNG-algebras: {\it observables SNG-algebras}

\n\be
\Obs^{\sS\sN\sG}  \m ,
\ee 
whether canonical or spacetime.
Each theory's  lattice of SNG generators or constraints  subalgebraic structures additionally 
induces a dual lattice of SNG observables                subalgebras. 
(If the reader is unsure what this means, they should consult the third Article of \cite{ALRoPoT} for the usual Lie subcase.)

\m 

\n 4) Deformation of SNG-algebraic structures encountering Rigidity \cite{G64, NR66, CM} gives a means of Constructing more structure from less.  
This generalizes, firstly, Spacetime Construction from space's `passing families of theories through the Dirac Algorithm' approach.   
Secondly, obtaining more structure from less for each of space and spacetime separately. 
(See e.g. Article IX in \cite{ALRoPoT} for Dirac and Lie Mathematics examples of each of these).
On the one hand, we can {\sl pose} rigidity for SNG-algebraic structures -- that under deformations of generators 

\n\be 
\scG \m \longrightarrow \m \scG_{\alpha}  \es  \scG + \alpha \, \phi   \m . 
\label{def}
\ee
for parameter $\alpha$ and functions $\phi$, the cohomology group condition

\n\be 
\mH^2(\Frg^{\sN\sS\sG}, \, \Frg^{\sN\sS\sG}) = 0    
\ee 
diagnoses rigidity.   
On the other hand, we are currently rather lacking in theorems for this SNG-algebra case.
The idea is moreover to take Rigidity to be \cite{Higher-Lie} a {\sl selection principle} in the Comparative Theory of Background Independence. 

\m 

\n 5) Reallocation of Intermediary-Object (RIO) Invariance gives the general SNG-algebraic structures' commuting-pentagon 
analogue of posing Refoliation Invariance for GR.
RIO  refers to whether, in going from an initial object to a final object, proceeding via intermediary object 1 or intermediary object 2 
causes one to be out by at most just an automorphism of the final object, 

\n\be 
O_{21}^{\sf\si\sn} - O_{12}^{\sf\si\sn} = Aut(O^{\sf\si\sn})  \m . 
\ee     
Refoliation Invariance is then the case in which our objects are spatial slices within a given spacetime. 
Consult Article III in \cite{ALRoPoT} for further details  about Refoliation Invariance, or \cite{Higher-Lie} and Article XII in \cite{ALRoPoT} for RIO Invariance more generally. 
This is again a {\sl selection principle} whose outcome remains unexplored in the SN case.  

\m 

\n 4) and 5)    are thus identifications of new research directions, 
whereas 1), 2) and 3) constitute             new confirmed structures and results.

\section{Conclusion}\label{Conclusion}

\n We have considered Nijenhuis Mathematics, with concrete focus on Schouten--Nijenhuis brackets and subsequent algebraic structures and algorithms. 
This is a useful generalization and robustness check -- a natural next port of call in developing the Comparative Theory of Background Independence -- 
and also a stepping stone to further cases of interest that follow from introduction of each of the following two further structural elements. 

\m 

\n A) The {\it Vinogradov bracket} \cite{V90, KS} $\mbox{\bf [} \m \mbox{\bf ,} \, \m \mbox{\bf ]}_{\sV}$ 
unifies the Schouten--Nijenhuis and Fr\"{o}licher--Nijenhuis brackets. 

\n B) The general {\it Gerstenhaber bracket} \cite{G63, Gengoux}, 
is a bilinear product taking one between spaces of $k$-linear maps $\mH\mC^k(V)$ of a graded vector space $\bFrV$: 

\n\be
\mbox{\bf [} \m \mbox{\bf ,} \, \m  \mbox{\bf ]}_{\sG} \m : \m \m  \mH\mC^{\bar{p}} \times \mH\mC^{\bar{q}} \longrightarrow \mH\mC^{\bar{p} + \bar{q}}  \m . 
\ee 
Like the SN bracket, this is of graded Lie bracket type, thus obeying graded-antisymmetry and graded-Jacobi identities similar to (2, 3).   
Such $k$-linear maps are moreover the basic objects entering {\it Hochschild cohomology} \cite{Gengoux, L98}, 
which then plays a major role in Deformation Quantization \cite{L78, Gengoux, Kontsevich} 
and the theory of quantum-mechanically relevant operator algebras \cite{Landsman}.  

\m 

\n A) and B) are known to support, firstly, {\it Vinogradov algebraic structures} and {\it Gerstenhaber algebraic structures}.  
Secondly -- as new results -- a {\it Vinogradov Algorithm} and a {\it Gerstenhaber Algorithm}: analogues of the Dirac and Lie Algorithms.   
Thirdly -- as new notions (as far as the author is aware) -- {\it Vinogradov} and {\it Gerstenhaber notions of observables}.
I.e.\ the zero commutants with the corresponding type of algebraic structure's generators under the corresponding defining type of bracket. 
Fourthly, one can moreover at least pose {\it Vinogradov} and {\it Gerstenhaber deformations}  
(not to be confused with Gerstenhaber having already worked out the deformations of simpler algebras). 
And then ask, as a selection principle, which subcases of each exhibit Rigidity.  
Finally, one can likewise pose {\it RIO Invariance in} both the {\it Vinogradov} and {\it general Gerstenhaber contexts}.   

\m 

\n{\bf Acknowledgments}  I thank various friends for support and proofreading.    


\end{document}